\UseRawInputEncoding
\documentclass[reprint,aps,prb,groupedaddress,showpacs,superscriptaddress]{revtex4-1}
\usepackage{graphicx}
\usepackage{dcolumn}
\usepackage[scheme=plain]{ctex}

\usepackage[normalem]{ulem}
\usepackage{color}

\usepackage{xcolor,cancel}

\begin{document}

\makeatletter
\newcommand{\rmnum}[1]{\romannumeral #1}
\newcommand{\Rmnum}[1]{\expandafter\@slowromancap\romannumeral #1@}
\makeatother

\title{Conformational change-modulated spin transport at the single-molecule level in carbon systems --Invited for the Third Carbon Special Topic}

\author{Yandong Guo}
\email{yandongguo@njupt.edu.cn}
\affiliation{College of Electronic and Optical Engineering, Nanjing University of Posts and Telecommunications, Nanjing 210046, China}
\affiliation{College of Natural Science, Nanjing University of Posts and Telecommunications, Nanjing 210046, China}
\affiliation{Key Laboratory of Radio Frequency and Micro-Nano Electronics of Jiangsu Province, Nanjing 210023, China}

\author{Xue Zhao}
\affiliation{College of Electronic and Optical Engineering, Nanjing University of Posts and Telecommunications, Nanjing 210046, China}

\author{Hongru Zhao}
\affiliation{College of Electronic and Optical Engineering, Nanjing University of Posts and Telecommunications, Nanjing 210046, China}

\author{Li Yang}
\affiliation{College of Electronic and Optical Engineering, Nanjing University of Posts and Telecommunications, Nanjing 210046, China}

\author{Liyan Lin}
\affiliation{College of Electronic and Optical Engineering, Nanjing University of Posts and Telecommunications, Nanjing 210046, China}
\affiliation{College of Natural Science, Nanjing University of Posts and Telecommunications, Nanjing 210046, China}

\author{Yue Jiang}
\affiliation{College of Electronic and Optical Engineering, Nanjing University of Posts and Telecommunications, Nanjing 210046, China}

\author{Dan Ma}
\affiliation{College of Electronic and Optical Engineering, Nanjing University of Posts and Telecommunications, Nanjing 210046, China}

\author{Yuting Chen}
\affiliation{College of Electronic and Optical Engineering, Nanjing University of Posts and Telecommunications, Nanjing 210046, China}

\author{Xiaohong Yan}
\affiliation{College of Electronic and Optical Engineering, Nanjing University of Posts and Telecommunications, Nanjing 210046, China}
\affiliation{Key Laboratory of Radio Frequency and Micro-Nano Electronics of Jiangsu Province, Nanjing 210023, China}
\affiliation{College of Science, Nanjing University of Aeronautics and Astronautics, Nanjing 210016, China}

\date{\today}

\begin{abstract}
Controlling the spin transport at the single-molecule level, especially without the use of ferromagnetic contacts, becomes a focus of research in spintronics. Inspired by the progress on atomic-level molecular synthesis, through first-principles calculations, we investigate the spin-dependent electronic transport of graphene nanoflakes with side-bonded functional groups, contacted by atomic carbon chain electrodes. It is found that, by rotating the functional group, the spin polarization of the transmission at the Fermi level could be switched between completely polarized and unpolarized states. Moreover, the transition between spin-up and spin-down polarized states can also be achieved, operating as a dual-spin filter. Further analysis shows that, it is the spin-dependent shift of density of states, caused by the rotation, that triggers the shift of transmission peaks, and then results in the variation of spin polarization. Such a feature is found to be robust to the {\color{black}{{length}}} of the nanoflake and the electrode material, showing great application potential. Those findings may throw light on the development of spintronic devices.

~\\
\noindent\textbf{Keywords}: spin-dependent electronic transport, molecular device, dual-spin filter, density-functional theory
\end{abstract}

\keywords{spin-dependent electronic transport, molecular device, dual-spin filter, density-functional theory}

\pacs{72.25.-b, 785.65.+h, 85.75.-d, 73.23.Ad}
\maketitle

\section{INTRODUCTION}
Recently, utilizing molecules to build electronic devices at the single-molecule level has been paid much attention, and it is considered to be a promising area for future nanoelectronics.\cite{thiele2014electrically,schedin2007detection,nozaki2010engineering,joachim1997electromechanical,selzer2006single}
Due to the development of techniques on selfassembly and nanofabrication, it is possible to fabricate practical molecular devices.\cite{tien1997microfabrication,xu1997nanometer}
Until now, kinds of functional molecular devices, as well as interesting electronic features, have been proposed, such as diode, logic operators, and negative differential resistance.\cite{chen2020conductance,wang2020theoretical,thiele2014electrically,dias2021investigation,niu2015phonon,min2021multifunctional,gu2018recent,li2021designing,song2021first,antonova2017negative,kobashi2017negative,rahighi2021all}
On the other hand, spintronics is also a rapidly emerging field for future devices, which exploiting the spin degree of freedom in electrons.\cite{hao2018spin,yang2019spin,peng2019electrically}
Thus, combining molecular electronics and spintronics together would no doubt offer more possibilities for device design.

Among all kinds of molecular structures, due to the peculiar atomic orbitals and bonding types, carbon-based ones exhibit various electronic behaviors and have attracted more and more attention, considered as a competitive candidate for next-generation electronic devices.\cite{stankovich2007synthesis,wang2010large,baughman2002carbon,howard1991fullerenes,zhang2013spin}
It has been reported that, with the tools of chemical synthesis, the ultimate goal of miniaturization in nanostructures' design can be arrived, where Chanteau et al.\cite{chanteau2003synthesis} successfully demonstrated the synthesization of 2-nm-tall carbon-based anthropomorphic molecules in monomeric, dimeric, and polymeric forms.
Interestingly, the anthropomorphic molecules could be synthesized in kinds of desired geometries, e.g., various heads (chef, jester, baker, and etc) and postures (dance, raising hands, holding hands, and etc).
This means the molecular synthesis can arrive at the atomic limit, providing us exciting opportunities for device design.

Both in theory and experiment, it is found that conformational change in molecules may induce the variation of
electronic behavior.
Especially, the twisting of carbon-based units could result in the variation of p-p coupling.\cite{venkataraman2006dependence,larsson1981electron,woitellier1989possibility,guo2013conformational,ma2010low,woitellier1989possibility} And in experiment, the rotation can be realized in many ways.\cite{tierney2011experimental,leoni2011controlling}
Based on those findings, kinds of functional device have been proposed, such as switch, amplifier and logic operators.
However, in these devices, the rotating parts mainly locate in the transport branch and play as the key bridge for electronic transmission. Such a kind of geometry usually limits the application scenarios of the device, as rotating the main part of the system may be not allowed or convenient.
Previous studies showed that, side-contacting molecules can also effectively influence the electronic transport of a nanosystem.\cite{chowdhury2011graphene}
According to those, together with the progress on atomic-level nanofabrication, it is expected to realize the modulation of electronic transport by rotating side-bonded functional groups in a single-molecule device, just like a valve in the pipeline. The study on it is still lacking, especially on spin-related features.
Such a device setup will not disturb the main structure of a configuration, showing advantages in manipulating electronic transport.

In the present work, we investigate the spin-dependent electronic transport of graphene nanoflakes with side-bonded functional groups, using the density functional theory (DFT) combined with nonequilibrium Green's function (NEGF).
{\color{black}{{ The two-probe systems are constructed by contacting the nanoflakes with atomic carbon chain electrodes. }}}
It is found that, the transmission at the Fermi level could be switched between completely polarized and unpolarized states, through rotating the functional group. Moreover, the transition between spin-up and spin-down polarized states can also be achieved, operating as a dual-spin filter. Further analysis shows that, it is the shift of density of states, caused by the rotation, that causes the shift of transmission peaks, and then results in the variation of spin polarization. And this is the intrinsic feature of this system, robust to the size of nanoflake and electrode material, indicating great application potential.

\begin{figure*}
 \includegraphics[width=1.0\textwidth]{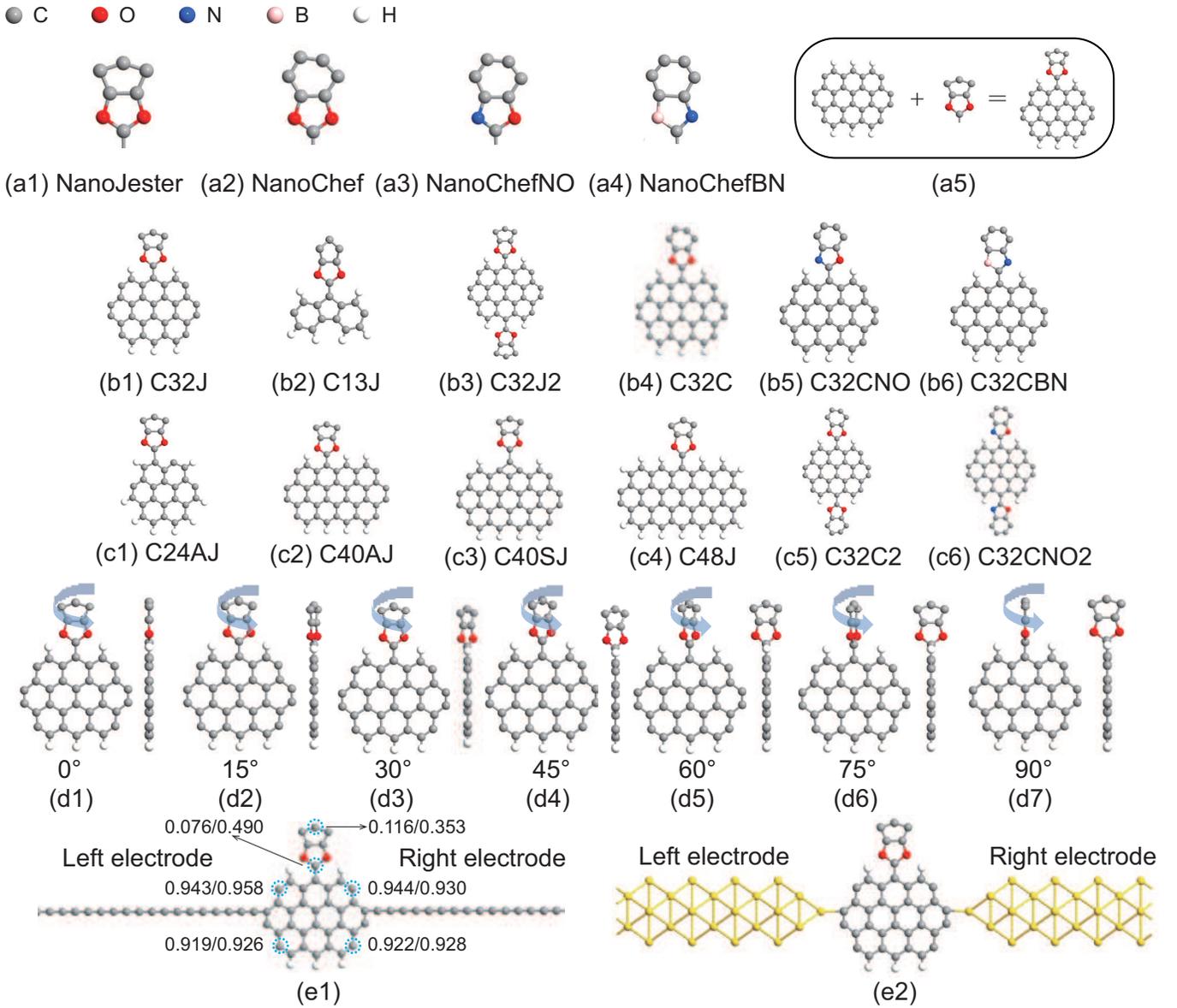}
 \caption{\label{structure}
(Color online) (a1)-(a4) The functional groups of NanoJester, NanoChef, NanoChefNO, and NanoChefBN, respectively. (a5) The illustration of combining graphene nanoflake and functional group to form C32J. (b1)-(b6) The geometries of C32J, C13J, C32J2, C32C, C32CNO, and C32CBN, respectively. (c1)-(c6) The geometries of C24AJ, C40AJ, C40SJ, C48J, C32C2, and C32CNO2, respectively. (d1)-(d7) The illustration of rotating the functional group in C32J. (e1)-(e2) The two-probe systems of C32J contacting with atomic carbon chain and Au nanowire electrodes, respectively. {\color{black}{{ The magnetic moments mainly locate on six carbon atoms, denoted by the dashed circles in (e1), and the corresponding magnetic moments are given out aside (the first and second values correspond to the magnetic moments of 0$^\circ$ and 90$^\circ$ rotation-angle cases respectively, where the unit of $\mu_{\text{B}}$ is omitted). }}}
}
\end{figure*}

\section{COMPUTATIONAL METHOD}
The calculations in the present work are carried out by combining DFT and NEGF together, which are implemented in the Atomistix Toolkit (ATK) package.\cite{taylor2001ab,brandbyge2002density,datta2000nanoscale,cohen2008insights} The electron exchange-correlation functional is treated by generalized gradient approximation (GGA) in the form of Perdew, Burke, and Ernzerhof (PBE).\cite{perdew1996generalized,perdew1992atoms,tao2003climbing} And the medium basis set of PseudoDojo pseudopotentials is used.\cite{van2018pseudodojo}
Sufficient vacuums (more than 10 {\AA}) in the supercell are constructed to eliminate the interactions between adjacent images.
The geometries are fully optimized until all the forces are less than 0.02 eV/{\AA}. The mesh cut-off energy of 150 Ry and 1$\times$1$\times$100 \emph{k}-point mesh in the Monkhorst-Pack scheme are employed.

\section{RESULTS AND DISCUSSIONS}
\begin{figure}
 \includegraphics[width=0.5\textwidth]{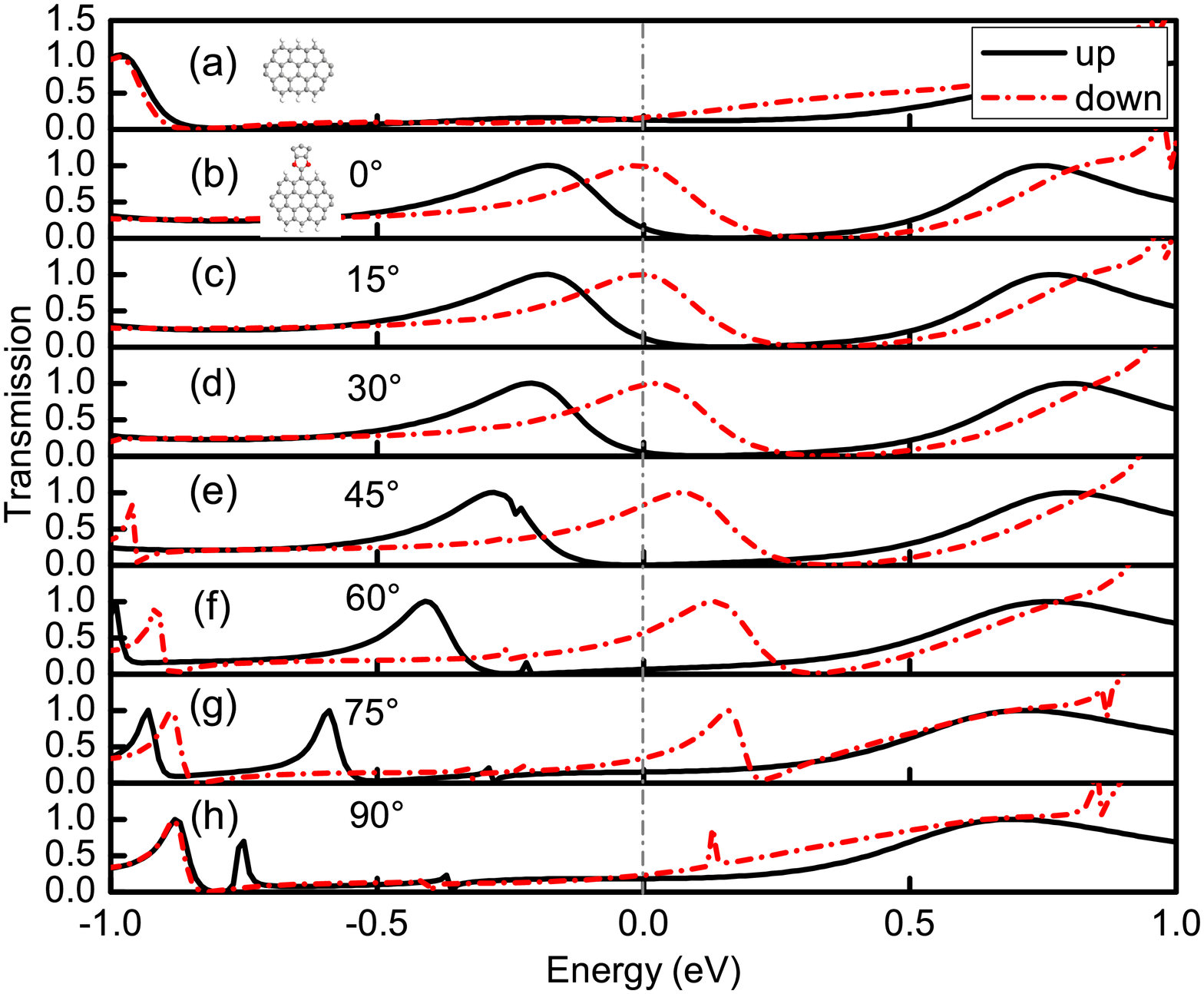}
 \caption{\label{C32J-trans}
(Color online) The transport spectra for two-probe systems that molecules contacted with atomic carbon chain electrodes. (a) for graphene nanoflake C32 (H is omitted for the configuration name). (b)-(h) for C32J with different rotation angles of the functional group of NanoJester, and the angle is denoted in each inset.
For the sake of simplicity, the electrodes are omitted in the figure, so as the following figures.
}
\end{figure}

\begin{figure}
 \includegraphics[width=0.5\textwidth]{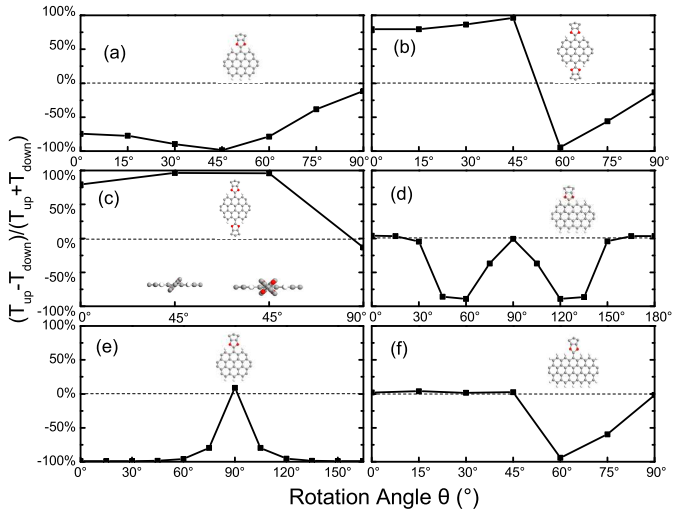}
 \caption{\label{SP-angle}
(Color online) The spin polarization at $E_F$ varies with the rotation angle for (a) C32J, (b) C32J2, (c) C32J2 , (d) C40AJ, (e) C32J (with Au electrodes), and (f) C48J systems, respectively. For C32J2 in (b), the two functional groups are rotated in the same direction, and for C32J2 in (c), they are rotated in the opposite directions.
}
\end{figure}

\begin{figure}
 \includegraphics[width=0.5\textwidth]{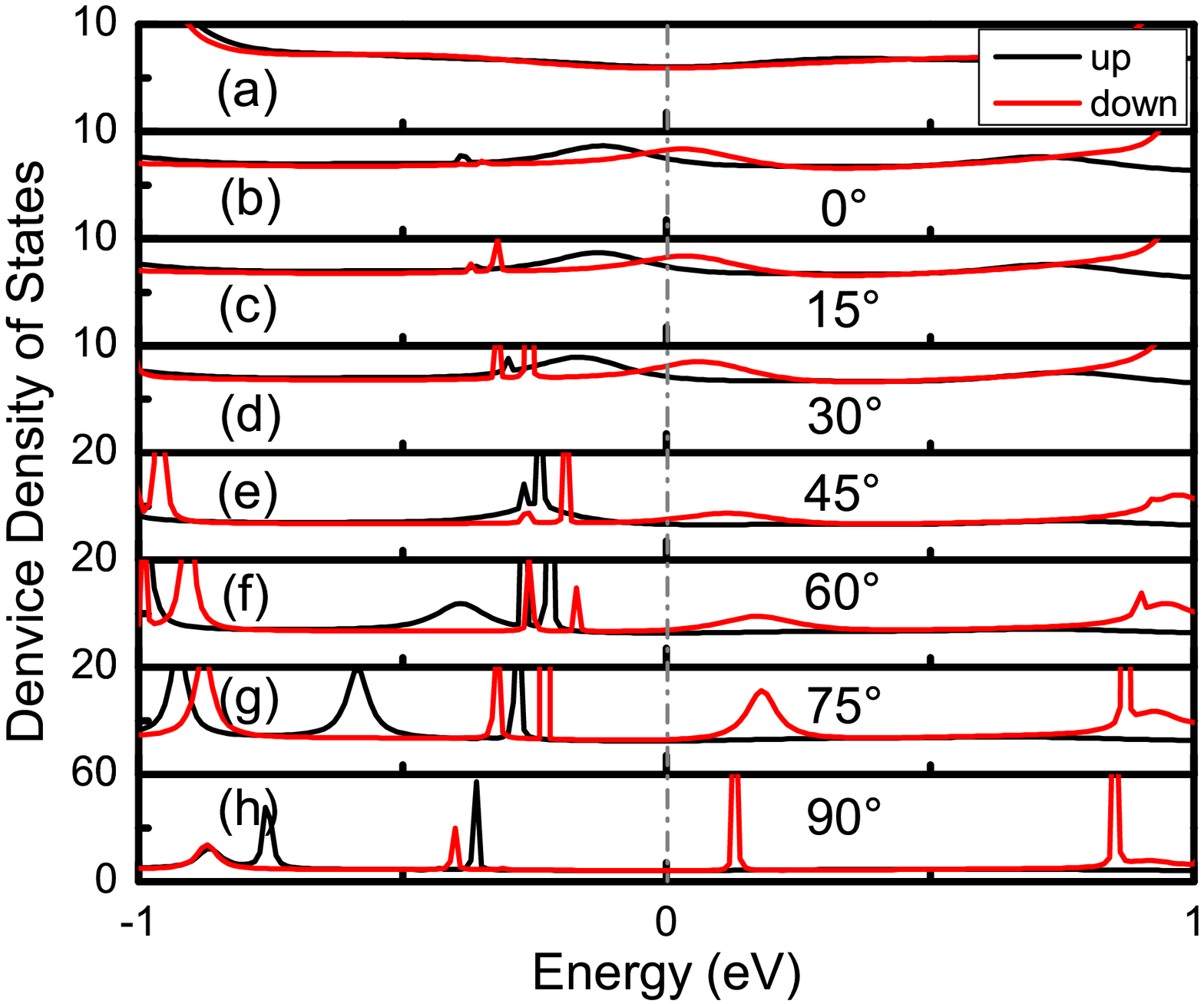}
 \caption{\label{DDOS-C32J}
(Color online) Device density of states for two-probe systems. (a) for graphene nanoflake C32. (b)-(h) for C32J with different rotation angles of NanoJester.
}
\end{figure}

Until now, there are kinds of carbon-based nanostructures being synthesized, e.g., graphene, nanotube, fullerene and etc. Among them, graphene is considered as one of the most promising candidates for future nanodevices. However, there is no spontaneous magnetism in it. Both theoretical and experimental investigations show that, cutting graphene to create edges, especially zigzag ones, is a feasible way to acquire spontaneous magnetic moment. With the development of miniaturization in electronics, the device at the single-molecule level is urgently needed. So, in the present work, we cut the graphene into nanoflakes with zigzag edges to explore the modulation of spin transport in it. To realize effective modulation, and at the same time to disturb the nanoflake as less as possible, we make carbon-based functional groups being bonded on the side of the flake.
In recent years, considering the fabrication and operation of a device, pure-carbon configurations attract increasing attention, as they would possess good stability.
Thus, we here choose atomic carbon chains as electrodes to contact with the nanoflakes.
The atomic carbon chain is a prefect kind of electrode with good conductivity.\cite{csahin2008first} And it could also form a good contact with the graphene nanoflake. This would help us to get rid of other electronic effects, and facilitate revealing the intrinsic feature of the system.

In carbon nanostructures, graphene is the most promising candidate for nanoelectronic device. To fulfill the miniaturization requirements, we cut graphene into nanoflakes. And to acquire spontaneous magnetism, graphene nanoflakes with zigzag edges are chosen.
Finally, a nanoflake denoted by C32 is constructed, shown in the left panel of Fig.~\ref{structure}(a5).
It can be seen as being cut from a zigzag graphene nanoribbon.
To mimic the fabrication process in experiment, the top and bottom edges of the nanoflake are hydrogenated. And to mimic the cutting process, the left and right edges are not hydrogenated, which can also facilitate the contact of electrodes.

{\color{black}{{
According to the experimental studies,\cite{chanteau2003synthesis} we here choose two kinds of functional groups, i.e., NanoJester and NanoChef, shown in Fig.~\ref{structure}(a1) and (a2) respectively.
Figure~\ref{structure}(a5) illustrates the combination of graphene nanoflake and NanoJester, and for clarity, the whole structure is denoted as C32J, shown in Fig.~\ref{structure}(b1).
The final two-probe setup of C32J is shown in Fig.~\ref{structure}(e1).
And Fig.~\ref{structure}(d1)-(d7) show how to rotate the functional group. In each figure, there is also a side view of the configuration.
}}}

\begin{figure}
 \includegraphics[width=0.5\textwidth]{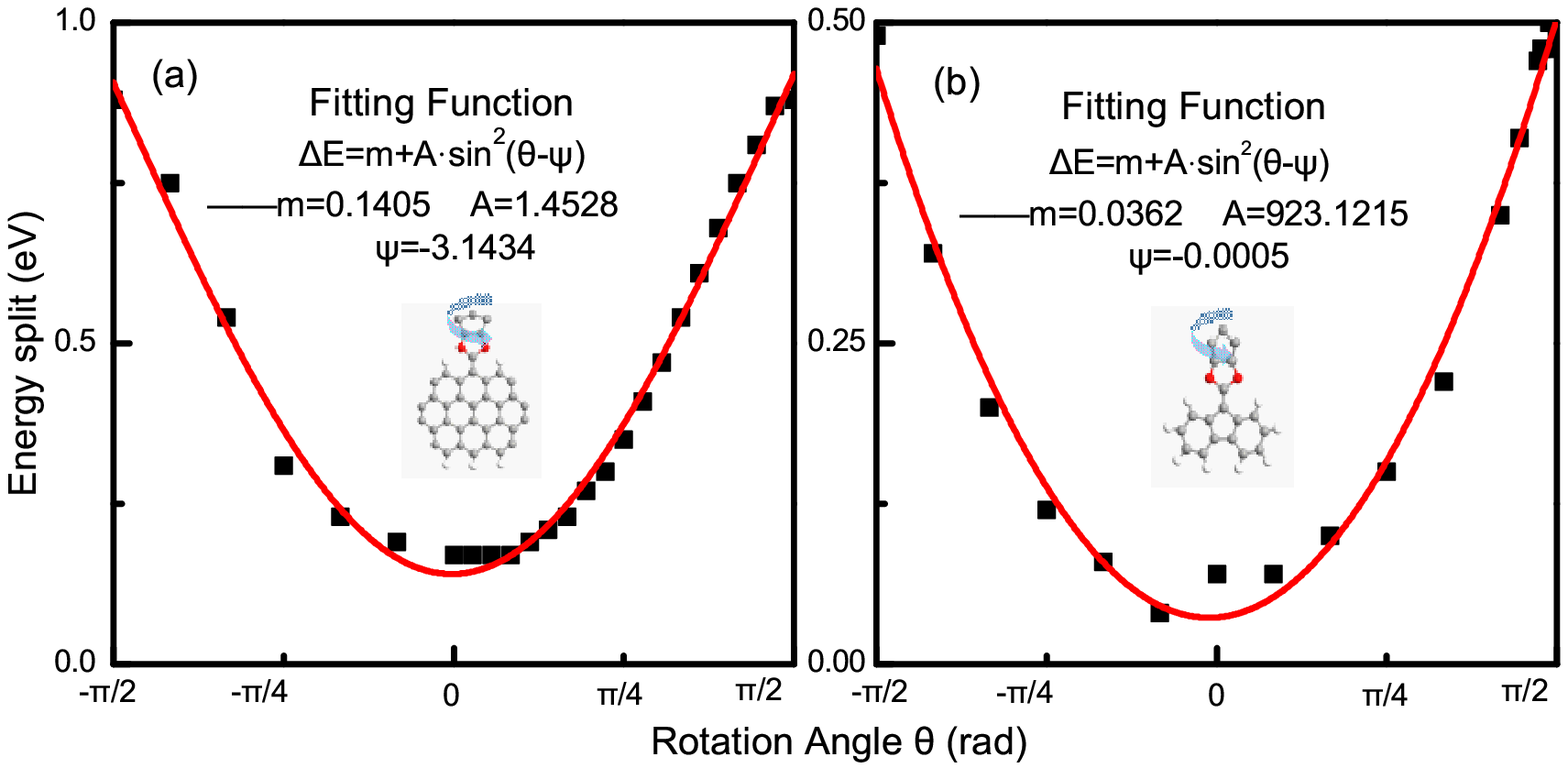}
 \caption{\label{Fitting}
(Color online) (a)-(b) Fitting functions for the energy split-angle relation in C32J and C13J systems, respectively. The energy split refers to the energy difference between spin-up and spin-down transmission peaks around $E_F$, see Fig.~\ref{C32J-trans} and Fig.~\ref{C13J-trans}.
The fitting function and parameters are present in the figure, and both of the two curves follow the sin$^2$($\theta$) relation.
}
\end{figure}

\begin{figure}
 \includegraphics[width=0.5\textwidth]{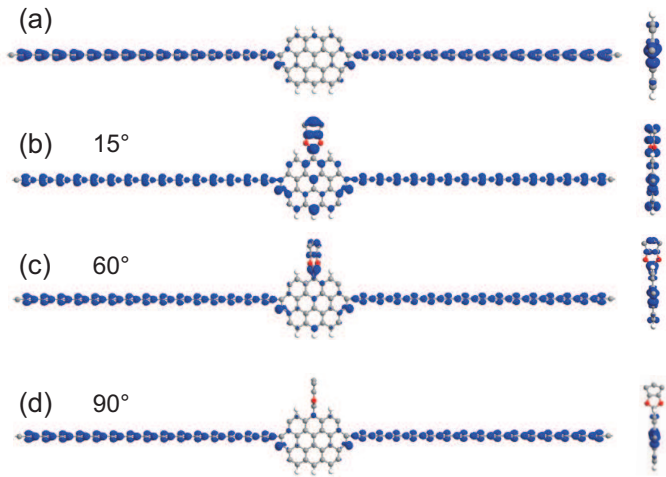}
 \caption{\label{LDDOS-C32J}
(Color online) Local density of states of the spin-down component at the Fermi level for two-probe systems. (a) for graphene nanoflake C32. (b)-(d) for C32J with rotation angles of 15$^\circ$, 60$^\circ$ and 90$^\circ$, respectively.
}
\end{figure}

\begin{figure}
 \includegraphics[width=0.5\textwidth]{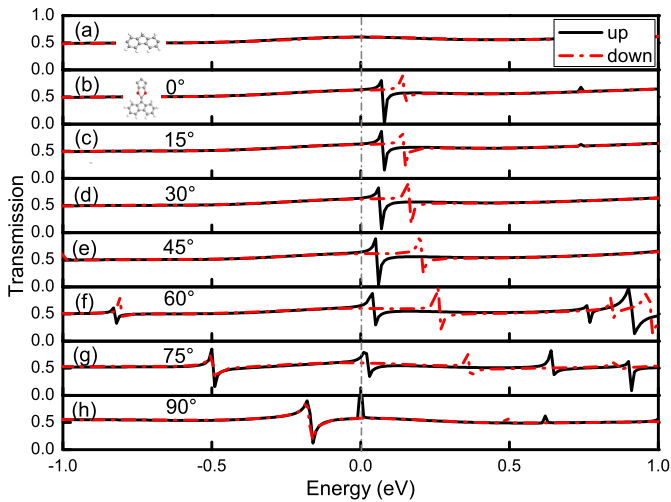}
 \caption{\label{C13J-trans}
(Color online) The transport spectra for molecules contacted with atomic carbon chain electrodes. (a) for graphene nanoflake C13. (b)-(h) for C13J with different rotation angles of NanoJester.
}
\end{figure}

\begin{figure}
 \includegraphics[width=0.5\textwidth]{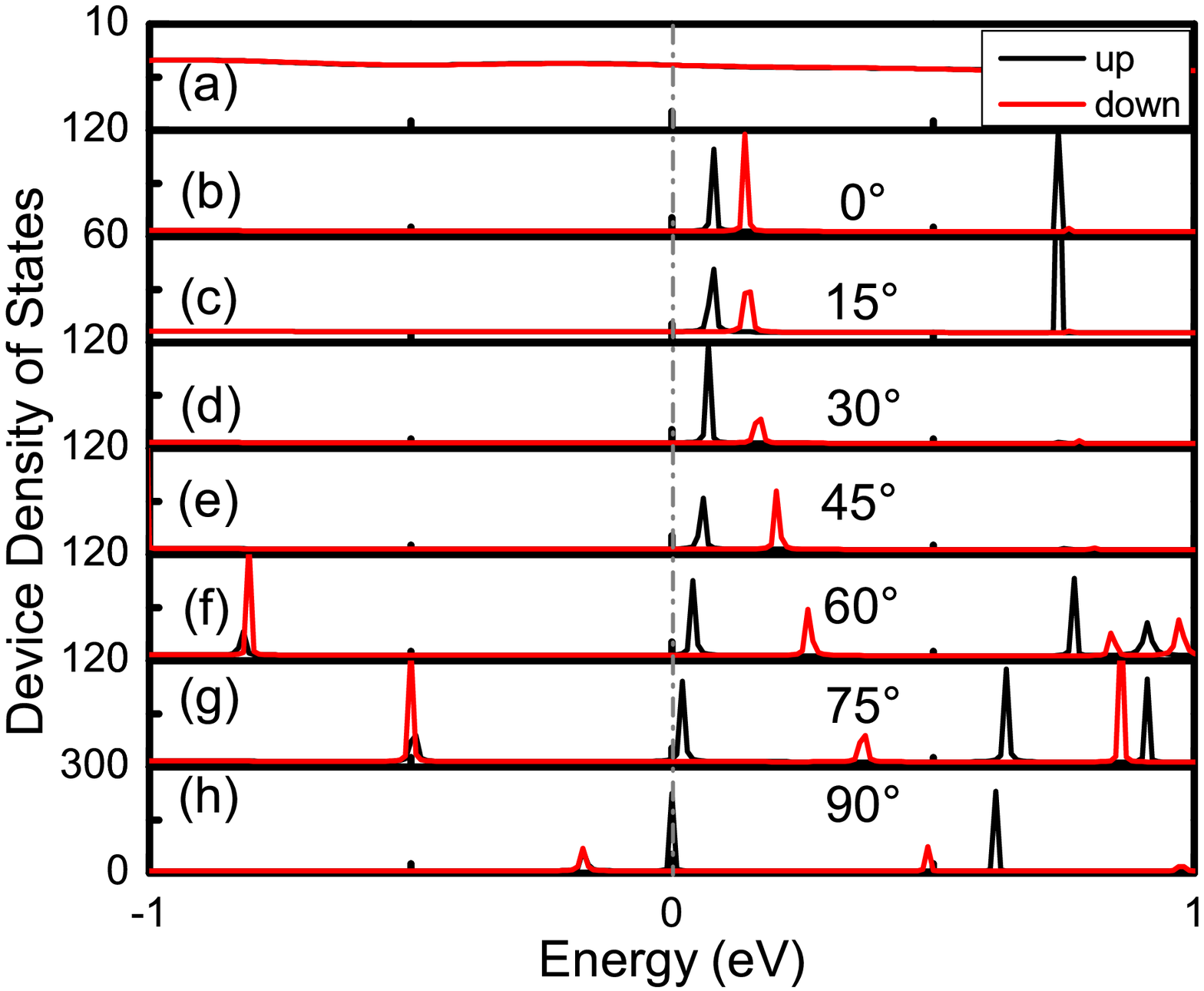}
 \caption{\label{DDOS-C13J}
(Color online) Device density of states for two-probe systems. (a) for graphene nanoflake C13. (b)-(h) for C13J with different rotation angles of NanoJester.
}
\end{figure}

To figure out the influence of the rotation on electronic transport, we calculate the spin-dependent transmission spectra for the two-probe system of C32J with different rotation angles, shown in Fig.~\ref{C32J-trans}(b)-(h). For the comparison with the bare nanoflake, the transmission spectra of C32 without the functional group is also calculated, shown in Fig.~\ref{C32J-trans}(a).
For the bare nanoflake, one finds the transport is finite and spin-unpolarized at $E_F$, see Fig.~\ref{C32J-trans}(a). Above $E_F$, the transport exhibits slightly spin-polarized. However, when the NanoJester is contacted with the nanoflake, the transport becomes quite different, see Fig.~\ref{C32J-trans}(b). One finds, around $E_F$, the transport is wholly increased. There are two transmission peaks emerges in this region, and they are in opposite spin components. More importantly, these two peaks are split in energy. And this results in a large spin-polarization at $E_F$, where the spin-down transmission reaches the summit and spin-up one is near the bottom of the valley.

Next, we rotate the functional group of C32J, and the rotation angle varies from 0$^\circ$ to 90$^\circ$. The corresponding transmission spectra are shown in Fig.~\ref{C32J-trans}(b)-(h). Interestingly, from the figures, one finds as the rotation angle increases, the spin-up transmission peak shifts left and spin-down one shifts right gradually.
Note that, the spin-up transmission peak is further away from $E_F$ than spin-down ones when $\theta=0^\circ$. As a result, when $\theta=45^\circ$, the spin-up peak completely shifts out of $E_F$, and spin-down one still locates there. So, the transmission becomes completely spin-polarized, operating as a spin filter. Moreover, the completely spin-polarized transmission distributes in a wide range around $E_F$, which is quite beneficial to the practical applications.

When $\theta>0^\circ$, the two peaks continue to shift in the opposite directions. However, at the same time, they both decrease gradually. So, at $E_F$, when $\theta=90^\circ$, the spin-down transmission goes down to a small value. Surprisingly, the spin-up transmission at $E_F$ increases when $\theta>45^\circ$, compared with that of $\theta=45^\circ$, and reaches an almost the same value with spin-down one when $\theta=90^\circ$. Thus, transmission becomes spin-unpolarized again at $E_F$, just like the non-NanoJester case in Fig.~\ref{C32J-trans}(a).
{\color{black}{{And in a wide range of energy around $E_F$, the transmission spectra are similar between non-NanoJester and $\theta=90^\circ$ cases, see Fig.~\ref{C32J-trans}(a) and (h).
But for the deep energy ranges above and below $E_F$, the transmission behaves different between $0^\circ$ and $90^\circ$.
Note that, for $\theta=90^\circ$, the planes of nanoflake and NanoJester are perpendicular to each other. So, such a $\theta=90^\circ$ effect may result from the destroy of the $\pi$-conjugation between nanoflake and NanoJester,\cite{venkataraman2006dependence,guo2013conformational} as well as the distribution of the magnetic moments (discussed below) and the spatial symmetry of the whole system.
}}}

{\color{black}{{
To observe the behavior of magnetism in the system, we do the Mulliken population analysis for the rotation angles of 0$^\circ$ and 90$^\circ$. It is found that, the magnetic moments mainly locate on six carbon atoms, and others are quite small. Those six atoms are denoted in Fig.~\ref{structure}(e1) and the corresponding magnetic moments are given out aside (the first and second values correspond to the magnetic moments of 0$^\circ$ and 90$^\circ$ cases respectively, where the unit of $\mu_{\text{B}}$ is omitted). One finds the magnetic moments of the four corner atoms in the nanoflake are around 0.94 $\mu_{\text{B}}$, which actually are the origin of the spin-dependent transport in the two-probe system. And they change quite a little from 0$^\circ$ to 90$^\circ$. However, for the two carbon atoms in the functional group, the magnetic moments change a lot from 0$^\circ$ to 90$^\circ$, suggesting modulation effect on the spin-related behaviors. For instance, the magnetic moment of the top carbon atom changes from 0.116 to 0.353 $\mu_{\text{B}}$ when the rotation angle varies from 0$^\circ$ to 90$^\circ$.
}}}

In the rotating process, one finds the spin-polarized state of transmission changes. To observe it more clearly, we plot the spin-polarization varying with the rotation angle, shown in Fig.~\ref{SP-angle}(a).
The spin-polarization here is defined as SP$=(T_{up}-T_{down})/(T_{up}+T_{down})$, and SP=0 means spin-unpolarized and SP=$\pm100\%$ means completely spin-polarized. For C32J in Fig.~\ref{SP-angle}(a), one finds SP could varies from almost 0 to -100\%. That is to say, through rotating, the modulation of spin polarization can be realized in a single-molecule system. Not only the completely polarized-unpolarized transition can be achieved, but also the state with any other SP value is expected to be achieved, just by finely tuning the rotation angle.
Those findings may bring many other novel devices.

{\color{black}{{
Apparently, it is the rotation-induced shift of the transmission peaks that causes the variation of spin polarization. }}}
To figure out the origin of the shift, we calculate and plot the device density of states (DDOS) of the two-probe C32J system, shown in Fig.~\ref{DDOS-C32J}, where (a)-(h) correspond to Fig.~\ref{SP-angle}(a)-(h) respectively.
From Fig.~\ref{DDOS-C32J}(a), one finds, for the non-NanoJester case, the DDOS is spin-unpolarized at $E_F$, the same as the transmission in Fig.~\ref{SP-angle}(a). However, when NanoJester is bonded, two DDOS peaks around $E_F$ emerge, see Fig.~\ref{DDOS-C32J}(b), and they locate at the same energy with the transmission peaks in Fig.~\ref{SP-angle}(b).
When rotating the NanoJester, the two DDOS peaks shift in the opposite directions, just like the transmission peaks, see Fig.~\ref{DDOS-C32J}(b)-(h). No doubt, it is those peak states in DDOS that contribute the transmission peaks around $E_F$, and so as the shift of them.
Although in some angle cases, e.g., 45$^\circ$-90$^\circ$, there are some small peaks arising around $E=-0.3$ eV, they are not delocalized enough to contribute electronic transmission and do not induce transmission peaks there.

As mentioned above, when the rotation angle increases, the two transmission peaks with opposite spins split and shift in the opposite directions. To see more clearly, we plot the variation of energy split between the two peaks following the angle $\theta$, shown in Fig.~\ref{Fitting}(a). The other rotating cases in the range of $\theta=[-\pi/2, 0]$ are also calculated and plotted. One finds, in the whole range of $\theta=[-\pi/2, \pi/2]$, the energy split does not change monotonously, and it decreases first and then increases with the increase of $\theta$. It reaches the minimum value when $\theta=0^\circ$.
Previous studies showed that, in a p-conjugated system, when the rotation angle varies, the p-p coupling strength will change following the square of trigonometric functions, e.g., cos$^2(\theta)$.\cite{woitellier1989possibility,larsson1981electron,guo2013conformational}
In our systems, although the variation of p-p coupling happens on the side of the nanoflake, there may also exists regular relationships between the angle and transmission behaviors.
Here, we try to fit those points with the square of a sine function, shown in Fig.~\ref{Fitting}(a).
{\color{black}{{
Obviously, one could find the curve follows the sin$^2(\theta)$ relation quite well.
This indicates that, in our systems, the rotation could effectively modulate the p-p coupling, which results in the shift of the orbitals. Such an effect that, a functional group's rotation-triggered orbital shift, has been observed experimentally in carbon-based systems.\cite{senge2000molecular} In our systems, the molecule is
contacted with two semi-infinite electrodes, so the molecular orbital will be broadened and contribute to the DDOS of the two-probe system. Consequently, besides the molecular states, the rotation would also induce the shift of DDOS, which can be characterized by the movement of the peaks.
For such a transport system, the transmission of electrons is mainly determined by the energy states of the middle molecule, which act like a bridge and contribute to the transmission channels.
Thus, the distribution of the states, i.e., DDOS, determines the behavior of the transport, and they (DDOS and transmission spectra) generally show a one-to-one correspondence. As a result, the rotation finally results in the shift of transmission spectra.
For our system, this shift effect is spin-dependent, where the spin-up and spin-down energy states may not shift synchronously, even in the opposite directions.
This asynchronous shift would result in the split of the DDOS peaks between spin-up and spin-down components, as well as the transmission ones.
As well known, the split of transmission peaks between different spins would trigger spin-polarized transport, which can be utilized to build spin devices, especially for the large spin polarization at the Fermi level.}}} Moreover, those finding may also throw light on the development of nanomechanics-related spintronics.

\begin{figure}
 \includegraphics[width=0.5\textwidth]{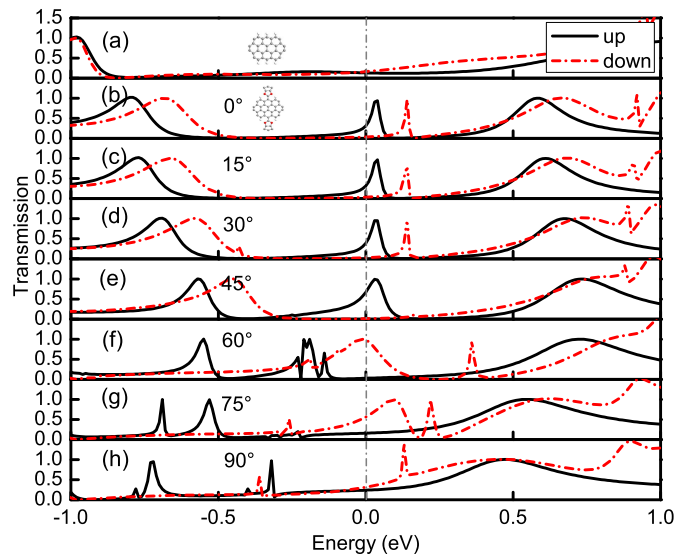}
 \caption{\label{trans-C32J2}
(Color online) The transport spectra for molecules contacted with atomic carbon chain electrodes. (a) for graphene nanoflake C32. (b)-(h) for C32J2 with different rotation angles of NanoJesters. The two NanoJesters are rotated in the same direction.
}
\end{figure}

\begin{figure}
 \includegraphics[width=0.5\textwidth]{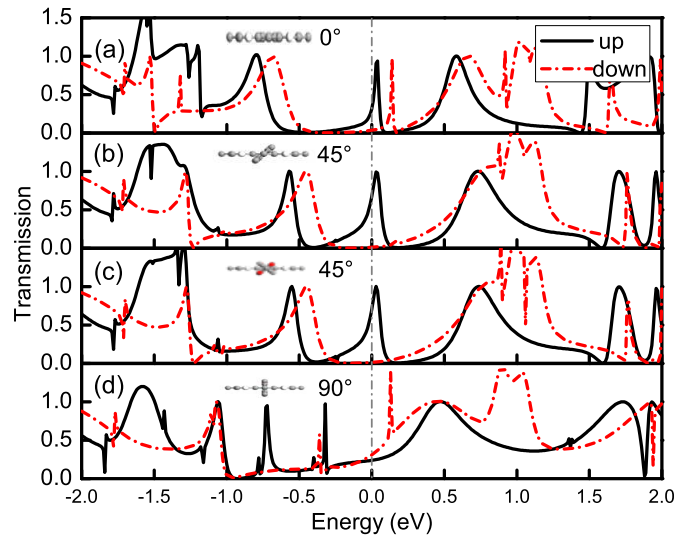}
 \caption{\label{trans-C32J2-45opposite}
(Color online) The transport spectra for molecules contacted with atomic carbon chain electrodes. (a) for graphene nanoflake C32. (b)-(d) for C32J2 with different rotation angles of NanoJesters, where the two NanoJesters are rotated in the same direction in (b) and in the opposite directions in (c).
}
\end{figure}

\begin{figure}
 \includegraphics[width=0.5\textwidth]{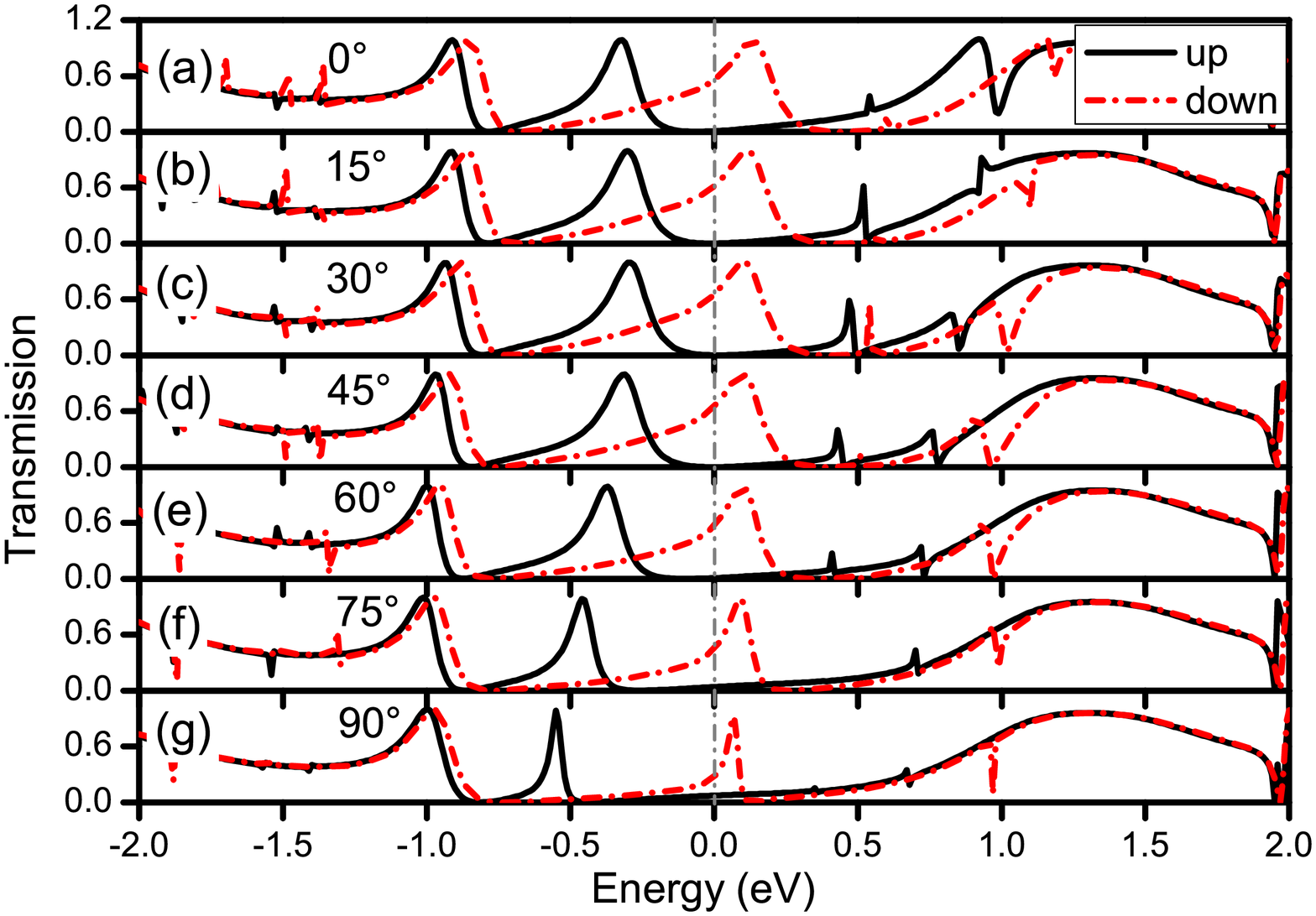}
 \caption{\label{C32CBN-trans}
(Color online) (a)-(g) The transport spectra for C32CBN systems with different rotation angles of NanoJesters, contacted by atomic carbon chain electrodes.
}
\end{figure}

\begin{figure}
 \includegraphics[width=0.5\textwidth]{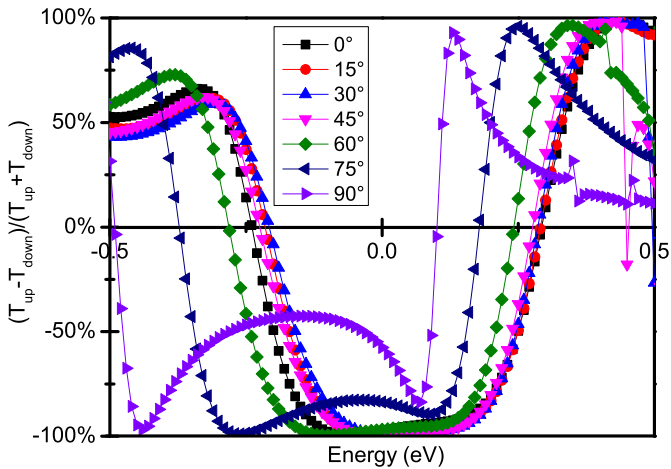}
 \caption{\label{C32CBN-SP}
(Color online) (a)-(g) The spin polarization for C32CBN systems with different rotation angles of NanoJesters, contacted by atomic carbon chain electrodes.
}
\end{figure}

\begin{figure}
 \includegraphics[width=0.5\textwidth]{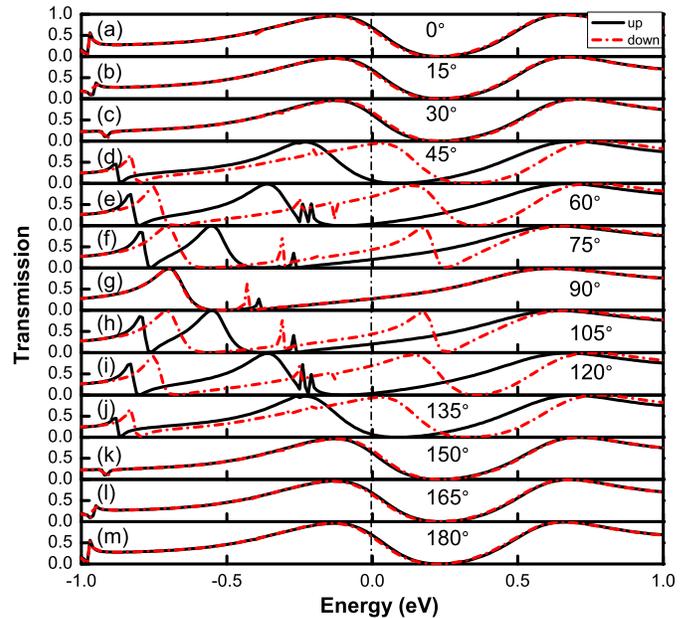}
 \caption{\label{C40AJ-trans}
(Color online) (a)-(m) The transport spectra for C40AJ systems with different rotation angles of NanoJesters, contacted by atomic carbon chain electrodes.
}
\end{figure}

As discussed in the above, it is the shift of the DDOS peaks that results in the variation of DDOS at $E_F$, which causes the variation of the transmission at $E_F$. To confirm and to see the variation process more clearly, we take the spin-down component as an example and calculate the local density of the states (LDOS) of the system at $E_F$ for different rotation angles, shown in Fig.~\ref{LDDOS-C32J}. Note that, for the sake of comparison, the isovalues are the same for all the cases in that figure.
One finds, for all the configurations, the LDOS is continuous on both left and right electrodes, suggesting good conductivity of the electrodes. The difference happens at the middle molecular region.
When there is no functional group, there is only a quite small distribution of LDOS, corresponding to the contributed small transmission at $E_F$ in Fig.~\ref{C32J-trans}(a).
When the NanoJester is bonded with $\theta=15^\circ$, the LDOS on the molecule increase a lot and become more continuous along the molecule, see Fig.~\ref{C32J-trans}(b). Thus, a large spin-down transmission emerges at $E_F$ in Fig.~\ref{C32J-trans}(c). When $\theta=60^\circ$, the LDOS decays quite a lot, compared with that of $\theta=15^\circ$, and the spin-down transmission also decreases, see Fig.~\ref{C32J-trans}(c) and Fig.~\ref{C32J-trans}(f).
However, when $\theta=90^\circ$, the LDOS at the molecule decays more severely, and consequently, the transmission decreases more heavily, see Fig.~\ref{C32J-trans}(d) and Fig.~\ref{C32J-trans}(h).
In brief, the spin-down transmission at $E_F$ decreases gradually with the rotation of the NanoJester. And one can easily see the functional group here operates like a valve in a pipeline.

Next, we change the molecule in the two-probe system to see weather the above transport features still exist.
Figure~\ref{structure}(b2) show the C13J configuration, which is constructed by combining a smaller graphene nanoflake and NanoJester. By contacting with atomic carbon chain electrodes, the transport spectra are calculated and shown in Fig.~\ref{C13J-trans}. In the figure, one can see there is also an energy split between spin-up and spin-down transmission peaks by bonding functional group, just a little above the Fermi level.
Moreover, by rotating the NanoJester, the split increases, the same as that of C32J system.
To trace the origin of the modulation, DDOS is calculated and plotted in Fig.~\ref{DDOS-C13J}. The same like that of C32J, one finds the DDOS peaks around $E_F$ shift in the opposite directions for different spins.
Obviously, it is those rotation-induced DDOS shift that results in the shift of transmission peaks.
Those results suggest such a rotation-modulated variation of energy split is a intrinsic feature of this kind of systems.
From the morphology of the peaks, traces of Fano resonance can be found, which is common in such kind of systems.\cite{csahin2008first,guo2010spin}
To check the effect of p-p coupling mechanism in this configuration, the fitting for the relation between rotation angle and energy split is also performed in the range of $\theta=[-\pi/2, \pi/2]$, shown in Fig.~\ref{Fitting}(b). Surprisingly, one finds the curve also follows the sin$^2(\theta)$ relation quite well.
Such kind of systems show good robustness to those transport features, which will be quite beneficial to practical applications.

{\color{black}{{
Note that, the transmission spectra of C13 and C13J with $\theta=90^\circ$ are different, even around the Fermi level, see Fig.~\ref{C13J-trans}(a) and (h) respectively. This is not like that of some other configurations, e.g., C32J, where the corresponding transmission spectra are similar, at least in the energy region around $E_F$. This indicates that, for such a C13 system, the width of the configuration might be an important factor. Actually, for the configuration of C13, it is more like a molecule, not a nanoflake, from a structural point of view. And for such a small molecule, each dimension, as well as each atom, plays an important role. While, for C32 and other configurations, they preserve the main geometric characters of graphene nanoribbons, e.g., two zigzag or armchair edges. To some extent, they can be seen as finite graphene nanoribbons. So, they are expected to inherit the electronic behaviors of graphene nanoribbons, where the edge morphology mainly dominates the electronic structure and transport, especially for the zigzag graphene nanoribbons. To figure out the influence of the width on different nanoflakes, a more detailed and systematic investigation is needed, which we hope to do in the future.
}}}

Next, we investigate the influence of the number of functional groups on the transport, and construct the configuration of C32J2, shown in Fig.~\ref{structure}(b3), where there are two NanoJesters bonded on the graphene nanoflake (top and bottom ones).
When rotating the two NanoJesters (in the same direction with the same angle), the transmission spectra are calculated and plotted in Fig.~\ref{trans-C32J2}.
This time, the energy split of the transmission peaks around $E_F$ also emerges.
Interestingly, when $\theta\leq45^\circ$, it is a spin-up transmission peak that dominates the transport at $E_F$, and the transmission is spin-up polarized (even near 100\% polarized for some cases). However, when $60^\circ\leq\theta\leq45^\circ$, it is a spin-down peak that dominates the transport there, and the transmission changes to spin-down polarized. And when $\theta=90^\circ$, the transport becomes almost spin-unpolarized. In brief, rotation could modulate the transport at $E_F$ switches among completely spin-up polarized, completely spin-down polarized, and unpolarized states.
This feature would be very helpful in device design.
To see the variation more clearly, we plot the spin polarization varying with $\theta$ at $E_F$, shown in Fig.~\ref{SP-angle}(b). From the figure, one could easily conclude that, other spin polarizations except the above three cases (0, 100\%, and -100\%) may also be achieved by finely tuning the rotation angle, e.g., SP=80\%.
So, a tunable dual-spin filter can be realized with our system.
This would provide many possibilities for device fabrication.

It should be noted that, there are two NanoJesters in the configuration. The rotations of the two functional groups could be in the opposite directions. As a demonstration, we calculate the transmission spectra for $\theta=45^\circ$ but in the opposite directions, shown in Fig.~\ref{trans-C32J2-45opposite}(c). For comparison, the transmission spectra of the configurations with $\theta=0^\circ, 45^\circ$ (in the same direction), $90^\circ$ are also presented, shown in Fig.~\ref{trans-C32J2-45opposite}(a), (b) and (d), respectively.
Apparently, the transmission spectra for the two kinds of $\theta=45^\circ$ cases are almost the same, especially around $E_F$. So, at least for those cases, the rotating directions of the two functional groups have little effect on the transport.

Beside the nanoflake, the functional group may also influence the electronic structure. So, we construct the functional groups of NanoChef, NanoChefNO and NanoChefBN, shown in Fig.~\ref{structure}(a2)-(a4) respectively, and also bond them to C32 nanoflake, shown in Fig.~\ref{structure}(b4)-(b6) respectively.
For the latter two functional groups, we change the two O atoms of (O, O) to (N, O) and (B, N) atoms, respectively. Different atoms could trigger electric polarization in the molecule, which would facilitate the rotation of the functional group by an electric field.
The transmission spectra of them under rotations are calculated and plotted.
However, for C32C and C32CNO, the transmission spectra are (almost) spin-unpolarized at $E_F$ and do not change with the variation of rotation angle, see Supporting Information. For C32CBN, the transmission spectra are completely spin-polarized around $E_F$ for a large energy range, and they are also insensitive to the rotation, see Fig.~\ref{C32CBN-trans}.
Such a large energy range with complete spin polarization is quite useful for spintronic device. To see it more clearly, we plot the spin polarization for all the angles, shown in Fig.~\ref{C32CBN-SP}.
One finds, when $\theta<75^\circ$, the transmission exhibits -100\% spin polarization for a wide energy around $E_F$, no matter how does the $\theta$ changes. This suggests good robustness, and this feature is suitable for practical applications.

To explore the effect of the number of NanoChef on the transport properties, we construct C32C2 and C32CNO2 configurations, shown in Fig.~\ref{structure}(c5) and (c6), respectively. The transmission spectra are show in Supporting Information. For them, there is no rotation-induced shift effect, where the transport is quite insensitive to the rotation. And there is no large spin-polarization around $E_F$, and even the transport is completely spin-unpolarized for some cases.

\begin{figure}
 \includegraphics[width=0.5\textwidth]{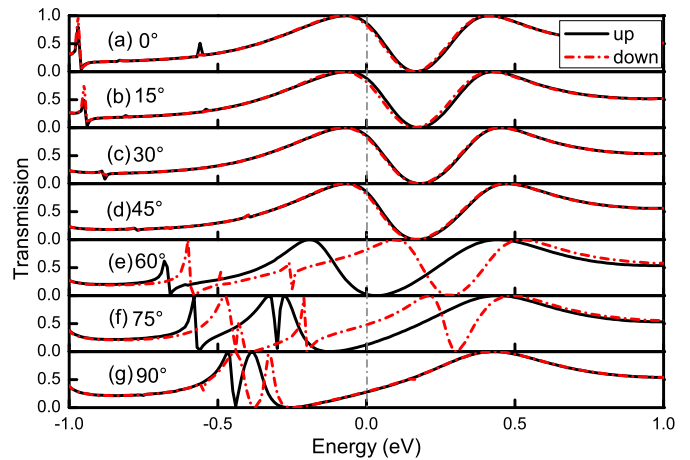}
 \caption{\label{C48J-trans}
(Color online) (a)-(g) The transport spectra for C48J systems with different rotation angles of NanoJesters, contacted by atomic carbon chain electrodes.
}
\end{figure}

\begin{figure}
 \includegraphics[width=0.5\textwidth]{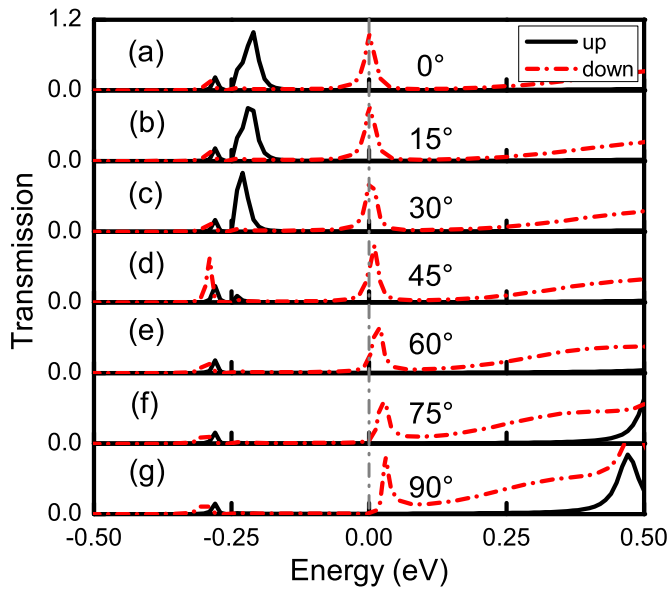}
 \caption{\label{C32J-Au-trans}
(Color online) (a)-(g) The transport spectra for C32J systems with different rotation angles of NanoJesters, contacted by Au nanowire electrodes.
}
\end{figure}

Next, we investigate the influence of the {\color{black}{{length}}} of the nanoflake on the electronic properties, and we construct C24AJ, C40AJ, C40SJ and C48J configurations, shown in Fig.~\ref{structure}(c1)-(c4), respectively, where S stands for symmetric and A stands for asymmetric.
For all these cases, the rotation-induced shift effect still exist.
For C24AJ and C40SJ configurations, the transmission spectra are almost spin-unpolarized or slightly spin-polarized at $E_F$, shown in Supporting Information. For C40AJ and C48J cases, the transmission spectra are shown in Fig.~\ref{C40AJ-trans} and Fig.~\ref{C48J-trans}, respectively. One finds, for both of the two cases, the transmission at $E_F$ could varies from almost 0 to -100\% by rotating. To see it more clearly, the SP varying with $\theta$ is shown in Fig.~\ref{SP-angle}(d) and (e), respectively.
Obviously, the SP could be modulated by rotation, and for C40AJ, there are two symmetric minus SP regions, although the configuration itself is not symmetric, see Fig.~\ref{SP-angle}(d) and Fig.~\ref{structure}(c2).

For nanodevice in applications, metal electrodes are also widely used. Among kinds of metal materials, Au electrodes possesses the best conductivity. To check the influence of metal electrodes on the transport, we take C32J as an example to contact with Au nanowire electrodes, shown in Fig.~\ref{structure}(e2). The transport spectra are shown in Fig.~\ref{C32J-Au-trans}. Surprisingly, the rotation-modulated shift effect remains, suggesting it is the intrinsic feature of those molecular systems and robust to the electrode material.
More importantly, the transmission at $E_F$ is completely spin-polarized when $\theta=0^\circ$. And due to the shift mechanism, the SP could reach to -100\% when rotating the NanoJester.
The variation of SP modulated by rotation is shown in Fig.~\ref{SP-angle}(f), and one can see a large range of SP=[0, -100\%] can be achieved. By projecting the DDOS onto the orbitals of different parts of the system, it is found that the peak around $E_F$ is mainly contributed by the p orbitals, especially the p orbitals on the middle molecule, i.e., C32J, see Supporting Information.
The realization of modulating SP in metal electrode systems could largely expand the application potential of the device.

{\color{black}{{
For practical applications, the stability of the structure is quite crucial. As for graphene flakes, plenty of theoretical and experimental investigations have been carried out on their stability.\cite{kuc2010structural,ricca2012infrared,wohner2014energetic,silva2010graphene,barnard2008thermal,ci2009graphene} From a geometric point of view, the configuration of C32, as well as C24, C40 and C48, can be seen as circular flake, and C13 can be seen as triangular flake. It is found that, those two kinds of graphene flakes exhibit good stability, especially after hydrogenation.\cite{kuc2010structural,ricca2012infrared,wohner2014energetic,silva2010graphene} For the functional groups we adopted, they have been synthesized in experiment and have been proved to be stable.\cite{chanteau2003synthesis} Moreover, edge functionalization on graphene flakes by such kind of functional groups has also been reported to be stable, both in theory and experiment.\cite{xiang2016edge,shao2020molecular,bellunato2016chemistry,sun2010soluble,dai2013functionalization}
}}}

\section{CONCLUSION}
In summary, through first-principles calculations, we investigate the spin-dependent electronic transport of
graphene nanoflakes with side-bonded functional groups, contacted by atomic carbon chain electrodes. It is found that, by rotating the functional groups, the spin polarization of the transmission at $E_F$ could be switched between completely spin-polarized and completely spin-unpolarized states. The transition between spin-up and spin-down polarized states can also be realized, operating as a dual-spin filter. Moreover, by tuning the rotation angle, any other partially spin-polarized state can be achieved. Further analysis shows that, it is the spin-dependent shift of DDOS peaks, caused by the rotation, that triggers the shift of transmission peaks, and then results in the variation of spin polarization at $E_F$. Such a feature is found to be robust to the {\color{black}{{length}}} of the nanoflake and the electrode material, showing great application potential.
Those findings may throw light on the fabrication of nanoelectronic devices.

\begin{acknowledgments}
This work is supported by the National Natural Science Foundation of China (11705097, 11504178 and 11804158), the Natural Science Foundation of Jiangsu Province (BK20170895), and the Funding of Jiangsu Innovation Program for Graduate Education (KYCX21$\_$0709).
\end{acknowledgments}

\bibliography{00-yl_head}

\end{document}